\documentclass[12pt]{article}

\author{Yu.~M.~Zinoviev
       \thanks{E-mail address: Yurii.Zinoviev@ihep.ru} \\
        {\it Institute for High Energy Physics} \\
        {\it Protvino, Moscow Region, 142280, Russia}}
\title{On massive spin 2 electromagnetic interactions}

\date{}

\begin{document}

\maketitle

\begin{abstract}
In this paper we investigate electromagnetic interactions for massive
spin 2 particles in $(A)dS$ space at linear approximation using gauge
invariant description for such massive particles. We follow bottom-up
approach, i.e. we begin with the introduction of minimal interaction
and then proceed by adding non-minimal interactions with higher and
higher number of derivatives together with corresponding non-minimal
corrections to gauge transformations until we are able to restore
gauge invariance broken by transition to gauge covariant derivatives.
We managed to construct a model that smoothly interpolates between
massless particle in $(A)dS$ space and massive one in a flat
Minkowski space. Also we reproduce the same results in a frame-like
formalism which can be more suitable for generalizations on higher
spins.
\end{abstract}

\thispagestyle{empty}
\newpage
\setcounter{page}{1}

\section*{Introduction}

It has been known since a long time that it is not possible to
construct standard gravitational interaction for massless higher spin
$s \ge 5/2$ particles in flat Minkowski space \cite{AD79,WF80,BBD85}
(see also recent discussion in \cite{Por08}). At the same time, it has
been shown \cite{FV87,FV87a} that this task indeed has a solution in
$(A)dS$ space with non-zero cosmological term. The reason is that
gauge invariance, that turns out to be broken when one replaces
ordinary partial derivatives by the gravitational covariant ones,
could be restored with the introduction of higher derivative
corrections containing gauge invariant Riemann tensor. These
corrections have coefficients proportional to inverse powers of
cosmological constant so that such theories do not have naive flat
limit. However it is perfectly possible, for cubic vertices, to have a
limit where both cosmological term and gravitational coupling constant
simultaneously go to zero in such a way that only interactions with
highest number of derivatives survive \cite{Zin08,BLS08}. Besides all,
it means that the procedure can be reversed. Namely, one can start
with the massless particle in flat Minkowski space and search for
non-trivial (i.e. with non-trivial corrections to gauge
transformations) higher derivatives cubic $s-s-2$ vertex containing
linearized Riemann tensor. Then, considering smooth deformation into
$(A)dS$ space, one can try to reproduce standard minimal gravitational
interaction as a by product of such deformation. Recently we have
shown that such procedure is indeed possible on the example of
massless spin 3 particle \cite{Zin08} using cubic four derivatives
$3-3-2$ vertex constructed in \cite{BL06} (see also \cite{BLS08} where
this vertex was reconsidered and an appropriate one for $s=4$ case has
been constructed).

Besides gravitational interaction one more classical and important
test for any higher spin theory is electromagnetic interaction. The
problem of switching on such interaction for massless higher spin
particles looks very similar to the problem with gravitational
interactions. Namely, if one replaces ordinary partial derivatives by
the gauge covariant ones the resulting Lagrangian loses its gauge
invariance and this non-invariance (arising due to non-commutativity
of covariant derivatives) is proportional  to field strength of vector
field. In this, for the massless fields with $s \ge 3/2$ in flat
Minkowski space  there is no possibility to restore gauge invariance
by adding non-minimal terms to Lagrangian and/or modifying gauge
transformations. But such restoration becomes possible if one goes to
$(A)dS$ space with non-zero cosmological constant. By the same reason,
as in the gravitational case, such theories do not have naive flat
limit, but it is possible to consider a limit where both cosmological
constant and electric charge simultaneously go to zero so that only
highest derivative non-minimal terms survive. Again it should be
possible to reproduce standard minimal e/m interaction starting with
some non-trivial cubic higher derivatives $s-s-1$ vertex containing
e/m field strength and considering its smooth deformation into $(A)dS$
space. An example of such procedure for massless spin 2 particle has
been given recently in \cite{Zin08a}, while candidate for appropriate
$s-s-1$ vertex was given in \cite{BLS08}.

In all investigations of massless particles interactions gauge
invariance plays a crucial role. Not only it determines a kinematic
structure of free theory and guarantees a right number of physical
degrees of freedom, but also to a large extent it fixes all possible
interactions of such particles. This leads, in particular, to
formulation of so-called constructive approach for investigation
of massless particles models 
\cite{OP65,FF79,MUF80,BBD85,Wald86,BH93,Hen98,BBCL06,BFPT06,BLS08}.
In this approach one starts with free Lagrangian for the collection of
massless fields with appropriate gauge transformations and tries to
construct interacting Lagrangian and modified gauge transformations
iteratively by the number of fields so that:
$$
{\cal L} \sim {\cal L}_0 + {\cal L}_1 + {\cal L}_2 + \dots, \qquad
\delta \sim \delta_0 + \delta_1 + \delta_2 + \dots
$$
where ${\cal L}_1$ --- cubic vertex, ${\cal L}_2$ --- quartic one and
so on, while $\delta_1$ --- corrections to gauge transformations
linear in fields, $\delta_2$ --- quadratic in fields and so on. 
In-particular, such approach allows one to consistently reproduce
such physically important theories as Yang-Mills, gravity and
supergravity.

It is natural to suggest that in any realistic higher spin theory
(like in superstring) most of higher spin particles must be massive
and their gauge symmetries spontaneously broken. But common
description of massive fields does not possess gauge invariance.
Instead, it requires that some constraints must follow from equations
of motion excluding all unphysical degrees of freedom. In this, at
least two general problems appear then one tries to switch on
interactions. First of all, a number of constraints could change thus
leading to a change in the number of degrees of freedom and
reappearing of unphysical ones. Secondly, even if a number of
constraints remains the same as in free theory, interacting theory
very often turns out to be non-causal, i.e. has solutions
corresponding to super-luminal propagation
\cite{VZ69,BKP99,BGKP99,BGP00,DPW00,DW01d}.
It is hard to formulate one simple principle for constructing
consistent theories with such particles. A number of different
requirements, such as conservation of right number of physical degrees
of freedom, smooth massless limit, tree level unitarity and causality,
was used in the past 
\cite{HS82,FPT92,CPD94,BKP99,BGKP99,BGP00,DPW00,DW01d}.

There exist two well known classes of consistent models for massive
high spin particles, namely, for massive non-Abelian spin 1 particles
and for massive spin 3/2 ones. In both cases masses of gauge fields
appear as a result of spontaneous gauge symmetry breaking. One of the
main ingredients of this mechanism is the appearance of Goldstone
particles with non-homogeneous gauge transformations. This, in turn,
leads to the gauge invariant description of such massive spin 1 and
spin 3/2 particles. But such gauge invariant description of massive
particles could be constructed for higher spins as well. There are at
least two basic approaches to such description. One of them is based
on the powerful BRST method \cite{BK05,BKRT06,BKL06,BKR07,MR07,BKT07}.
Another one appeared as an attempt to generalize to higher spins a 
mechanism of spontaneous gauge symmetry breaking
\cite{Zin83,KZ97,Zin01,Met06} (see also
\cite{Zin02a,Zin03a,Med03,BHR05,HW05,BG08}). In such a breaking a set
of Goldstone fields with non-homogeneous gauge transformations appear
making gauge invariant description of massive gauge fields possible. 
Such gauge invariant description of massive fields works well not only
in flat Minkowski space-time, but in (anti) de Sitter space-times as
well. All that one needs to do is to replace ordinary partial
derivatives with the covariant ones and take into account commutator
of these derivatives which is non-zero now. In particular, this
formulation turns out to be very convenient for investigation of 
so-called partially massless theories which appear in de Sitter space
\cite{DW01,DW01a,DW01c,Zin01,SV06,DW06}. The mere existence of gauge
invariant formulation for massive higher spin particles allows us to
extend the constructive approach for any collection of massive and/or
massless particles, see e.g. \cite{KZ97,Zin06,Met06a,Zin08}.

In a gauge invariant formalism the problem of switching on
gravitational or electromagnetic interactions for massive particles
looks very similar to that for the massless ones. Namely, introduction
of minimal interactions by the replacement of ordinary partial
derivatives by the covariant ones spoils the invariance of the
Lagrangian under gauge transformations. Having at our disposal mass
$m$ as a dimensionfull parameter even in a flat Minkowski space we can
try to restore broken gauge invariance by adding to the Lagrangian
non-minimal terms containing the linearized Riemann tensor (e/m field
strength) as well as corresponding non-minimal corrections to gauge
transformations. Naturally such terms will have coefficients
proportional to inverse powers of mass $m$ so that the theory will not
have naive massless limit. However, it is natural to suggest that
there exists a limit where both mass and gravitational coupling
constant (electric charge) simultaneously go to zero so that only some
interactions containing Riemann tensor (e/m field strength) survive.
In this, an interesting and important question is the relation between
flat space limit for massless particles in $(A)dS$ space and massless
limit for massive particles in flat Minkowski space. To understand
such relation (if any) it is important to consider general case ---
massive particles in $(A)dS$ space with arbitrary cosmological
constant.

The first step towards the construction of gravitational interactions
for massive spin 3 particles was performed in \cite{Zin08}, while in
the Section 1 we give simple but illustrative example of
electromagnetic interactions for massive spin 3/2 particles.
The main purpose of this paper is to investigate electromagnetic
interaction for massive spin 2 particles in $(A)dS$ space. In Section
2 we begin with the metric-like formulation where the main gauge field
is a symmetric second rank tensor $h_{(\mu\nu)}$, while vector $B_\mu$
and scalar $\varphi$ fields play the role of Goldstone ones. We follow
the bottom-up approach, i.e. we begin with the introduction of minimal
interaction and then add non-minimal terms with higher and higher
number of derivatives until we are able to restore broken gauge
invariance. The first such possibility arises when we add to the
Lagrangian terms with two derivatives as well as one derivative
corrections to gauge transformations. Such a model gives a
generalization of our previous results \cite{KZ97} to the case of
arbitrary $(A)dS$ space. But it turns out that for any non-zero value
of cosmological constant such model is singular in the massless limit.
So we proceed with three derivatives vertices in the Lagrangian and
two derivatives corrections to gauge transformations. Among all
solutions there is one unambiguous model having non-singular massless
limit. In this model we obtain the following relation for the electric
charge, mass and cosmological constant:
$$
e_0 = - a_0 [ m^2- \kappa(d-2)] \frac{d-3}{d-2}
$$
where $a_0$ --- coupling constant for main three derivatives vertex
having dimension $1/m^2$. Let us stress that that the main three
derivatives $2-2-1$ vertex is exactly the same as in \cite{Zin08a}, so
such model indeed smoothly interpolates between massless particle in
$(A)dS$ space and massive one in a flat Minkowski space.

As is well known, basically there are two approaches for description
of gravity theory --- metric one, where the main object is symmetric
metric tensor $g_{\mu\nu}$, and tetrad one with tetrad $e_\mu{}^a$ and
Lorentz connection $\omega_\mu{}^{ab}$. These two approaches admit
natural generalization for description of higher spin particles.
Generalization of metric approach has been constructed in
\cite{SH74,SH74a,Fro78,FF78,Fro79,FF80}, while generalization of
tetrad approach, the so-called frame-like formalism, has been
constructed in \cite{Vas80,LV88,Vas88} (see also
\cite{AD80,Zin03,Zin03a,ASV03,ASV05,SV06,ASV06,SV08,Skv08,Zin08b,
Zin08c,BIS08,BIS08a}). In Section 3  we reproduce the results of
previous section using frame-like gauge invariant formulation for
massive particles in $(A)dS$ space \cite{Zin03a,Zin08b}, The main
reason is that frame-like formulation being elegant and geometric in
nature could be more suitable for generalizations on higher spins.

\section{Example with spin 3/2}

In this section as a simple but instructive example we consider
electromagnetic interactions of massive spin 3/2 particle (see also
\cite{DPW00,DW01d}). We will work in general $(A)dS_4$ space with
arbitrary cosmological constant and use the following conventions on
$(A)dS$ covariant derivatives acting on spinors:
\begin{equation}
[ D_\mu, D_\nu ] \eta = - \frac{\kappa}{2} \sigma_{\mu\nu} \eta,
\qquad \kappa = \frac{2\Lambda}{(d-1)(d-2)} = \frac{\Lambda}{3},
\qquad \sigma_{\mu\nu} = \frac{1}{2} [ \gamma_\mu, \gamma_\nu ]
\end{equation}
To construct a gauge invariant description for massive spin 3/2
particle we need vector-spinor $\Psi_\mu$ as well as spinor $\chi$. It
is easy to check that the following Lagrangian:
\begin{equation}
{\cal L}_0 = \frac{i}{2} \varepsilon^{\mu\nu\alpha\beta}
\bar{\Psi}_\mu \gamma_5 \gamma_\nu D_\alpha \Psi_\beta + \frac{i}{2}
\bar{\chi} \hat{D} \chi + \frac{M}{2} \bar{\Psi}_\mu
\sigma^{\mu\nu} \Psi_\nu + i \sqrt{\frac{3}{2}} m (\bar{\Psi} \gamma)
\chi + M \bar{\chi} \chi
\end{equation}
where $\hat{D} = \gamma^\mu D_\mu$, $M = \sqrt{m^2 - \kappa}$, is
invariant under the following local gauge transformations:
\begin{equation}
\delta_0 \Psi_\mu = D_\mu \eta - \frac{i M}{2} \gamma_\mu \eta
\qquad \delta_0 \chi = \sqrt{\frac{3}{2}} m \eta
\end{equation}
Recall that in a de Sitter space ($\kappa > 0$) we have unitary
forbidden region $m^2 < \kappa$ (see e.g. \cite{Gar03}).

Now let us introduce minimal electromagnetic interaction. We prefer to
work with Majorana spinors so in what follows we will assume that all
spinor objects are doublets: 
$$
\Psi_\mu = \left( \begin{array}{c} \Psi_\mu{}^1 \\ \Psi_\mu{}^2
\end{array} \right), \qquad
\chi = \left( \begin{array}{c} \chi^1 \\ \chi^2 \end{array} \right),
\qquad
\eta = \left( \begin{array}{c} \eta^1 \\ \eta^2 \end{array} \right) 
$$
Thus we replace $(A)dS$ covariant derivatives by the fully covariant
ones:
\begin{equation}
D_\mu \Rightarrow \nabla_\mu = D_\mu + e_0 q A_\mu, \qquad
q = \left( \begin{array}{cc} 0 & 1 \\ -1 & 0 \end{array} \right),
\qquad q^2 = - I
\end{equation}
As usual such replacement breaks the invariance of the Lagrangian
under the local gauge transformations and we obtain:
\begin{equation}
\delta_0 {\cal L}_0 = i e_0 \bar{\Psi}_\mu q \tilde{F}^{\mu\nu}
\gamma_5 \gamma_\nu \eta, \qquad \tilde{F}^{\mu\nu} = \frac{1}{2}
\varepsilon^{\mu\nu\alpha\beta} F_{\alpha\beta}
\end{equation}
So we will try to restore broken invariance (at least in the linear
approximation) by adding to the Lagrangian the most general
non-minimal terms, containing electromagnetic field strength:
\begin{eqnarray}
{\cal L}_1 &=& \frac{1}{2} \bar{\Psi}_\mu \left[ a_1 F^{\mu\nu} +
a_2 \gamma_5 \tilde{F}^{\mu\nu} + a_3 g^{\mu\nu} (\sigma F) + a_4
(F^{\mu\alpha} \sigma_\alpha{}^\nu + \sigma^{\mu\alpha}
F_\alpha{}^\nu) \right] q \Psi_\nu + \nonumber \\
 && + i \bar{\Psi}_\mu ( a_5 F^{\mu\nu} + a_6 \gamma_5
\tilde{F}^{\mu\nu}) \gamma_\nu q \chi + \frac{a_7}{2} \bar{\chi} q 
(\sigma F) \chi
\end{eqnarray}
as well as the most general corrections to gauge transformations:
\begin{eqnarray}
\delta_1 \Psi_\mu &=& i q (\alpha_1 F_{\mu\nu} + \alpha_2 \gamma_5
\tilde{F}_{\mu\nu}) \gamma^\nu \eta \qquad \delta_1 \chi = q
\alpha_3 (\sigma F) \eta \nonumber \\
\delta_1 A_\mu &=& \alpha_4 (\bar{\Psi}_\mu q \eta) + i \alpha_5
(\bar{\chi} \gamma_\mu q \eta)
\end{eqnarray}
First of all we calculate all variations with two derivatives and
require their cancellation. Simple calculations give:
$$
\alpha_2 = \alpha_1, \qquad \alpha_4 = 2 \alpha_1, \qquad
\alpha_5 = - 2 \alpha_3
$$
$$
a_1 = - a_2 = - 2 \alpha_1, \qquad a_3 = a_4 = 0, \qquad
a_5 = a_6 = - 2 \alpha_3
$$
In this, non-minimal Lagrangian and appropriate corrections to gauge
transformations take the form familiar from supergravity models:
\begin{equation}
{\cal L}_1 = - \alpha_1 \bar{\Psi}_\mu (F^{\mu\nu} - \gamma_5
\tilde{F}^{\mu\nu}) q \Psi_\nu + i \alpha_3 \bar{\Psi}_\mu (\sigma F)
\gamma^\mu q \chi  + \frac{a_7}{2} \bar{\chi} q (\sigma F) \chi
\end{equation}
\begin{eqnarray}
\delta_1 \Psi_\mu &=& - \frac{i \alpha_1}{2} q (\sigma F) \gamma_\mu
\eta \qquad \delta_1 \chi = q \alpha_3 (\sigma F) \eta \nonumber \\
\delta_1 A_\mu &=& 2 \alpha_1 (\bar{\Psi}_\mu q \eta) - 2i \alpha_3
(\bar{\chi} \gamma_\mu q \eta)
\end{eqnarray}
At last cancellation of variations with one derivative (taking into
account term coming from the introduction of minimal interactions)
gives:
$$
e_0 = 2 \alpha_1 M + 2 \sqrt{6} \alpha_3 m, \qquad
a_7 = - \frac{4 M}{\sqrt{6} m} \alpha_3
$$
A few comments are in order.
\begin{itemize}
\item If we calculate a commutator of two gauge transformations we
obtain e.g.:
\begin{equation}
[ \delta_1, \delta_2] A_\mu = - 4 i (\alpha_1{}^2 + 2
\alpha_3{}^2) (\bar{\eta_2} \gamma^\nu \eta_1) F_{\nu\mu}
\end{equation}
This means that for non-zero value of electric charge $e_0$ any such
model must be a part of some (spontaneously broken) supergravity
theory.
\item From the supergravity point of view the meaning of two
parameters $\alpha_1$ and $\alpha_3$ is clear: in the most general
case our vector field can be a linear combination of graviphoton (with
vector-spinor $\Psi_\mu$ as a superpartner) and some vector field from
vector supermultiplet (with spinor superpartner).
\item From the expression for the parameter $a_7$ above, one can see
that in general there is an ambiguity between massless and flat
limits (see also \cite{DW00}). Indeed, in the flat Minkowski space we
obtain $a_7 = - 4 \alpha_3 /\sqrt{6}$ and nothing prevents us from
considering massless limit $m \to 0$. But for any non-zero
cosmological constant the expression for $a_7$ is singular in the
massless limit. \item The most simple model free from such ambiguity
is the case $\alpha_3 = 0$, i.e. our photon is a pure graviphoton. In
this case an effective electric charge is given by $e_0 = 2 \alpha_1
M$ so that it becomes equal to zero exactly at the boundary of
unitary allowed region.
\end{itemize}

\section{Metric-like formalism}

In this section we consider electromagnetic interaction for massive
spin 2 particles in $(A)dS_d$ space with arbitrary cosmological
constant using metric-like gauge invariant formalism \cite{Zin01}. We
need three fields: symmetric second rank tensor $h_{\mu\nu}$, vector
$B_\mu$ and scalar $\varphi$ ones. As is well known, even for massless
spin 2 particles in $(A)dS$ space gauge invariance requires
introduction of mass-like terms into Lagrangian. So in what follows we
will organize the calculations just by the number of derivatives. For
example, gauge invariant Lagrangian for free massive spin 2 particle
will be written as follows:
$$
{\cal L}_0 = {\cal L}_{02} + {\cal L}_{01} + {\cal L}_{00}
$$
where first index '0' means free (quadratic in fields) theory, while
the second one denotes a number of derivatives. Note also that due to
non-commutativity of $(A)dS$ covariant derivatives there is some
ambiguity in the structure of kinetic terms for massless spin 2
particle. We will use the following concrete choice:
\begin{eqnarray}
{\cal L}_{02} &=& \frac{1}{2} D^\alpha h^{\mu\nu} D_\alpha h_{\mu\nu}
- \frac{1}{2} D^\alpha h^{\mu\nu} D_\mu h_{\nu\alpha} - \frac{1}{2}
(D h)^\mu (D h)_\mu + (D h)^\mu D_\mu h - \frac{1}{2} D^\mu h D_\mu h
- \nonumber \\
 && - \frac{1}{2} (D_\mu B_\nu - D_\nu B_\mu)^2 + \frac{2(d-1)}{d-2}
(D_\mu \varphi)^2 \\
{\cal L}_{01} &=& 2m (h^{\mu\nu} D_\mu B_\nu - h (D B)) + 
\frac{4(d-1)M}{d-2} (D B)\varphi \\
 {\cal L}_{00} &=& - \frac{M^2}{2} ( h^{\mu\nu} h_{\mu\nu} - h^2 )
- \frac{2(d-1)mM}{d-2} h \varphi + \frac{2d(d-1)m^2}{(d-2)^2}
\varphi^2 - 2\kappa(d-1) B_\mu{}^2
\end{eqnarray}
where $M^2 = m^2 - \kappa(d-2)$. Recall that in de Sitter space
($\kappa > 0$) we again have unitary forbidden region $m^2 < \kappa
(d-2)$. Similarly, the gauge transformations leaving this Lagrangian
invariant will be written as follows
$$
\delta_0 = \delta_{01} + \delta_{00}
$$
\begin{equation}
\delta_{01} h_{\mu\nu} = D_\mu \xi_\nu + D_\nu \xi_\mu, \qquad
\delta_{01} B_\mu = D_\mu \lambda
\end{equation}
\begin{equation}
\delta_{00} h_{\mu\nu} = \frac{2m}{d-2} g_{\mu\nu} \lambda, \qquad
\delta_{00} B_\mu = m \xi_\mu, \qquad
\delta_{00} \varphi = M \lambda
\end{equation}
where again first index '0' means initial (non-homogeneous) gauge
transformations, while the second one denotes a number of derivatives.
As for the $(A)dS$ covariant derivatives in this section we will use
the following normalization:
\begin{equation}
 [ D_\mu, D_\nu ] v_\alpha = R_{\mu\nu,\alpha\beta} v^\beta = - \kappa
( g_{\mu\alpha} v_\nu - g_{\nu\alpha} v_\mu ), \qquad
\kappa = \frac{2 \Lambda}{(d-1)(d-2)}
\end{equation}

Recall that one of the nice features of gauge invariant formulation
for massive fields is that it admits a smooth massless limit. Indeed,
if we consider the limit $m \to 0$ for non-zero value of cosmological
constant the total Lagrangian decomposes into the sum of free
Lagrangians describing massless spin 2 and massive spin 1 particles
(or into the sum of massless Lagrangians for spin 2, 1 and 0 particles
in flat case). In this, total number of physical degrees of freedom
remains  the same as in massive case. Note that working with such
description one is often used to "eliminate" additional fields by
simply setting them to 0. Such procedure may be useful as a simple and
quick way to check the number of degrees of freedom and we will use
it in the Conclusion to show that interacting model constructed
in this paper does have correct number of degrees of freedom. Let us
stress however that such simplified procedure does not "commute" with
taking massless limit. Indeed, if we simply set vector and scalar
fields to 0 and then consider massless limit we will get massless spin
2 theory without any trace of other degrees of freedom. As usual in
any gauge invariant theory, the rigorous way consists of complete
analysis of all first class constraints and appropriate gauge fixing. 

Now let us introduce minimal electromagnetic interaction. First of all
we add to our Lagrangian usual kinetic terms for e/m field:
$$
{\cal L}_0 \quad \Rightarrow \quad {\cal L}_0 - 
\frac{1}{4} F_{\mu\nu}{}^2
$$
 We prefer to
work with real fields so we will assume that all our fields are real
doublets $h_{\mu\nu}{}^i$, $B_\mu{}^i$ and $\varphi^i$ where $i=1,2$.
Thus we replace all derivatives in the Lagrangian and gauge
transformations by fully covariant ones:
$$
D_\mu \xi_\nu{}^i \rightarrow D_\mu \xi_\nu{}^i - e_0
\varepsilon^{ij} A_\mu \xi_\nu{}^j
$$
As usual such replacement spoils the invariance of the Lagrangian
under gauge transformations:
\begin{eqnarray}
\delta_0 {\cal L}_0 &=& e_0 \varepsilon^{ij} [ - 3
F_{\mu\nu} D_\mu h_{\nu\alpha}{}^i - 3 (D h)_\mu{}^i F_{\mu\alpha} + 3
D_\mu h^i F_{\mu\alpha} - h_{\mu\nu}{}^i D_\mu F_{\nu\alpha} - 
(D F)_\mu h_{\mu\alpha}{}^i + \nonumber \\
 && \qquad + (D F)_\alpha h^i - 4 m B_\mu{}^i F_{\mu\alpha} ]
\xi_\alpha{}^j + e_0 \varepsilon^{ij} B_{\mu\nu}{}^i
F_{\mu\nu} \lambda^j \label{res0}
\end{eqnarray}
where $F_{\mu\nu} = D_\mu A_\nu - D_\nu A_\mu$.

So we will try to restore broken gauge invariance by adding
non-minimal terms containing e/m field strength $F_{\mu\nu}$ to the
Lagrangian as well as appropriate corrections to gauge
transformations. The simplest possibility is to add all possible terms
with one derivative:
\begin{equation}
{\cal L}_{11} = \varepsilon^{ij} F_{\mu\nu} [ a_1
h_{\mu\alpha}{}^i h_{\nu\alpha}{}^j + a_2 B_\mu{}^i B_\nu{}^j ]
\label{lag1}
\end{equation}
as well as the most general corrections to gauge transformations
without derivatives:
\begin{equation}
\delta_{10} A_\mu = \varepsilon^{ij} [ \alpha_1 h_{\mu\nu}{}^i
\xi_\nu{}^j + \alpha_2 h^i \xi_\mu{}^j + \alpha_3 \varphi^i
\xi_\mu{}^j + \alpha_4 B_\mu{}^i \lambda^j ] \label{cor1}
\end{equation}
but it could be easily checked that it is impossible to achieve gauge
invariance by adjusting parameters $a_{1,2}$ and $\alpha_{1,2,3,4}$.

Thus we proceed by adding all possible two derivatives terms to the
Lagrangian:
\begin{eqnarray}
{\cal L}_{12} &=& \varepsilon^{ij} F^{\mu\nu} [ b_1 D_\mu 
h_{\nu\alpha}{}^i B_\alpha{}^j + b_2 (D h)_\mu{}^i B_\nu{}^j + b_3
D_\mu h^i B_\nu{}^j + \nonumber \\
 && \qquad \quad + b_4 h^i B_{\mu\nu}{}^j + b_5 h_{\mu\alpha}{}^i
D_\nu B_\alpha{}^j + b_6 h_{\mu\alpha}{}^i D_\alpha B_\nu{}^j +
\nonumber \\
 && \qquad \quad + b_7 D_\mu \varphi^i B_\nu{}^j + b_8 B_{\mu\nu}{}^i
\varphi^j ] \label{lag2}
\end{eqnarray}
where $B_{\mu\nu}{}^i = D_\mu B_\nu{}^i - D_\nu B_\mu{}^i$,
as well as the most general corrections to gauge transformations with
one derivative:
\begin{eqnarray}
\delta_{11} B_\mu{}^i &=& \beta_0 \varepsilon^{ij} F_{\mu\nu}
\xi_\nu{}^j \nonumber \\
\delta_{11} A_\mu &=& \varepsilon^{ij} [ \beta_1 D_\mu B_\nu{}^i
\xi_\nu{}^j + \beta_2 D_\nu B_\mu{}^i \xi_\nu{}^j + \beta_3 (D B)^i
\xi_\mu{}^j + \nonumber \\
 && \qquad + \beta_4 B_\nu{}^i D_\nu \xi_\mu{}^j + \beta_5 B_\nu{}^i
D_\mu \xi_\nu{}^j + \beta_6 B_\mu{}^i (D \xi)^j + \nonumber \\
 && \qquad + \rho_1 (D h)_\mu{}^i \lambda^j + \rho_2 D_\mu h^i
\lambda^j + \rho_3 D_\mu \varphi^i \lambda^j + \nonumber \\
 && \qquad + \rho_4 h_{\mu\nu}{}^i D_\nu \lambda^j + \rho_5 h^i D_\nu
\lambda^j + \rho_6 \varphi^i D_\mu \lambda^j ] \label{cor2}
\end{eqnarray}
Note that due to gauge invariance of the free Lagrangian gauge
transformations in this linear approximation are defined up to
possible field dependent free gauge transformations
$\delta A_\mu \sim D_\mu X$ only. In other words, gauge
transformations are always defined up to possible redefinitions of
gauge parameters. In linear approximation for massless fields the
choice made does not change anything, though for massive fields the
structure of gauge transformations for Goldstone fields does depend
on the choice made. In what follows we will always use all possible
redefinitions of gauge parameters to bring gauge transformations to as
simple form as possible. Here we will use this ambiguity to set
$\beta_5 = \rho_2 = \rho_3 = 0$. Also recall that any
interaction Lagrangian where the number of derivatives is greater or
equal to that of free Lagrangian is always determined up to possible
field redefinitions. For the case at hands such redefinitions have the
form:
$$
A_\mu \Rightarrow A_\mu + \varepsilon^{ij} [ \kappa_1
h_{\mu\nu}{}^i B_\nu{}^j + \kappa_2 h^i B_\mu{}^j + \kappa_3 \varphi^i
B_\mu{}^j ]
$$
In what follows we choose $\rho_4 = \rho_5 = \rho_6 = 0$. First of all
we consider variations with three derivatives and require their
cancellation:
$$
\delta_{01} {\cal L}_{12} + \delta_{11} {\cal L}_{02} = 0
$$
This allows us to express all parameters $b$, $\beta$ and $\rho$ in
terms of one main parameter $\beta_0$:
$$
b_1 = b_2 = b_3 = 0, \qquad
b_4 = - \frac{\beta_0}{2}, \qquad b_5 = - 2 \beta_0, \qquad 
b_6 = 2 \beta_0, \qquad b_7 = 0
$$
$$
\beta_1 = - \beta_2 = 2 \beta_0, \qquad \beta_3 = \beta_4 = 0,
\qquad \rho_1 = 0
$$
Then we add terms with one derivative (\ref{lag1}) to the Lagrangian
as well as corrections without derivatives (\ref{cor1}) to the gauge
transformations and require cancellation of variations with two and
one derivative:
$$
\delta_{01} {\cal L}_{11} + \delta_{00} {\cal L}_{12} +
\delta_{11} {\cal L}_{01} + \delta_{10} {\cal L}_{02} = 0
$$
$$
\delta_{00} {\cal L}_{11} + \delta_{11} {\cal L}_{00} +
\delta_{10} {\cal L}_{01} = 0
$$
taking into account terms (\ref{res0}) coming from the
introduction of minimal e/m interactions. We obtain:
$$
\beta_0 = - \frac{e_0}{m}, \qquad b_8 = \frac{2(d-1)e_0 M}{(d-2)m^2}
$$
$$
a_1 = \frac{e_0}{2}, \qquad 2a_2 = - \alpha_4 = 
\frac{4e_0 (m^2 - \kappa(d-1))}{m^2}
$$
$$
\alpha_1 = - 2e_0, \qquad \alpha_2 = 0, \qquad \alpha_3 = 2 b_8 m
$$
Collecting all results we see that gauge invariance broken by the
introduction of minimal e/m interaction could be restored (in linear
approximation) with the introduction of the following non-minimal
terms:
\begin{eqnarray}
{\cal L}_1 &=& - \frac{2e_0}{m} \varepsilon^{ij} [ h_{\mu\nu}{}^i
B_{\mu\alpha}{}^j F_{\nu\alpha} - \frac{1}{4} h^i B_{\mu\nu}{}^j
F_{\mu\nu} ] + b_8 \varepsilon^{ij} \varphi^i B_{\mu\nu}{}^j
F_{\mu\nu} + \nonumber \\
 && + \varepsilon^{ij} F_{\mu\nu} [ \frac{e_0}{2}
h_{\mu\alpha}{}^i h_{\nu\alpha}{}^j + a_2 B_\mu{}^i B_\nu{}^j ]
\end{eqnarray}
supplemented with the following corrections for gauge transformations:
\begin{eqnarray}
\delta_1 B_\mu{}^i &=& - \frac{e_0}{m} \varepsilon^{ij} F_{\mu\nu}
\xi_\nu{}^j \nonumber \\
\delta_1 A_\mu &=&  \varepsilon^{ij} [ - \frac{2e_0}{m} B_{\mu\nu}{}^i
\xi_\nu{}^j  - 2e_0 h_{\mu\nu}{}^i \xi_\nu{}^j  + 2b_8 m \varphi^i
\xi_\mu{}^j + \alpha_4 B_\mu{}^i \lambda^j ]
\end{eqnarray}
In the flat space limit ($\kappa = 0$) these results agree (up to
slightly different field normalization) with our previous results in
\cite{KZ97} thus providing their generalization into $(A)dS$ space.
But from the expressions for the parameters $b_8$, $a_2$ and
$\alpha_4$ above one can see that for any non-zero value of
cosmological term $\kappa$ such model is singular in the limit $m \to
0$, $e_0 \to 0$, $e_0/m = const$. Besides flat space limit there is
only one non-singular case corresponding to so-called partially
massless spin 2 particles \cite{DW01,DW01a,DW01c,Zin01}. Indeed, if
one put $m^2 = \kappa (d-2)$ the scalar fields $\varphi^i$ completely
decouple. The free Lagrangian for such particle has the form:
\begin{eqnarray}
{\cal L}_0 &=& \frac{1}{2} [ D_\mu h_{\alpha\beta}{}^i D_\mu
h_{\alpha\beta}{}^i - (D h)_\mu{}^i (D h)_\mu{}^i - D_\mu
h_{\alpha\beta}{}^i \partial_\alpha h_{\mu\beta}{}^i + 2 (D h)_\mu{}^i
D_\mu h^i - D_\mu h^i D_\mu h^i ] - \nonumber \\
 && - \frac{1}{2} B_{\mu\nu}{}^i B_{\mu\nu}{}^i 
 + 2 m [ h_{\mu\nu}{}^i D_\mu B_\nu{}^i - h^i (D B)^i ] -
2 \kappa (d-1) B_\mu{}^i B_\mu{}^i 
\end{eqnarray}
being invariant under the following gauge transformations:
\begin{equation}
\delta_0 h_{\mu\nu}{}^i = D_\mu \xi_\nu{}^i + D_\nu \xi_\mu{}^i +
\frac{2m}{d-2} g_{\mu\nu} \lambda^i, \qquad
\delta_0 B_\mu{}^i = D_\mu \lambda^i + m \xi_\mu{}^i
\end{equation}
In this, non-minimal interactions which are necessary to restore gauge
invariance after introduction of minimal e/m interaction look like:
\begin{equation}
{\cal L}_1 = - \frac{2e_0}{m} \varepsilon^{ij} [ h_{\mu\nu}{}^i
B_{\mu\alpha}{}^j F_{\nu\alpha} - \frac{1}{4} h^i B_{\mu\nu}{}^j
F_{\mu\nu} ] + \varepsilon^{ij} F_{\mu\nu} [ \frac{e_0}{2}
h_{\mu\alpha}{}^i h_{\nu\alpha}{}^j - \frac{2 e_0}{d-2} B_\mu{}^i
B_\nu{}^j ]
\end{equation}
while appropriate corrections to gauge transformations have the form:
\begin{equation}
\delta_1 B_\mu{}^i = - \frac{e_0}{m} \varepsilon^{ij} F_{\mu\nu}
\xi_\nu{}^j, \qquad
\delta_1 A_\mu =  \varepsilon^{ij} [ - \frac{2e_0}{m} B_{\mu\nu}{}^i
\xi_\nu{}^j  - e_0 h_{\mu\nu}{}^i \xi_\nu{}^j  
+ \frac{4 e_0}{d-2} B_\mu{}^i \lambda^j ]
\end{equation}

Let us return back to the general case --- massive theory in $(A)dS$
space with arbitrary cosmological term. As we have recently shown
\cite{Zin08a} to obtain e/m interactions for massless spin 2
particles in $AdS$ space one needs non-minimal interactions with three
derivatives. So it seems natural to suppose that to construct massive
theory having non-singular limit $m \to 0$ one has to consider all
possible corrections up to three derivatives as well. 

We begin with the three derivatives vertex that played crucial role
for the massless theory \cite{Zin08a}:
\begin{eqnarray}
{\cal L}_{13} &=& a_0 \varepsilon_{ij} F^{\mu\nu} [ - D_\mu 
h_{\alpha\beta}{}^i D_\alpha h_{\beta\nu}{}^j - \frac{1}{2}
D_\alpha h_{\beta\mu}{}^i D_\alpha h_{\beta\nu}{}^j
+ D_\alpha h_{\beta\mu}{}^i D_\beta h_{\alpha\nu}{}^j +
\nonumber \\
 && \qquad \quad + \frac{1}{2} D_\mu h_{\alpha\beta}{}^i
D_\nu h_{\alpha\beta}{}^j - D_\mu h_{\nu\alpha}{}^i
(D h)_\alpha{}^j -  \frac{1}{2} (D h)_\mu{}^i (D h)_\nu{}^j +
\nonumber \\
 && \qquad \quad + (D h)_\mu{}^i D_\nu h^j +
D_\mu h_{\nu\alpha}{}^i D_\alpha h^j - \frac{1}{2}
D_\mu h^i D_\nu h^j ]
\end{eqnarray}
together with appropriate corrections to gauge transformations:
\begin{eqnarray}
\delta_{12} h_{\mu\nu}{}^i &=& a_0 \varepsilon^{ij} [ \frac{1}{2} (
F_{\mu\alpha} D_{[\alpha} \xi_{\nu]}{}^j + F_{\nu\alpha} 
D_{[\alpha} \xi_{\mu]}{}^j ) + \frac{1}{d-2} g_{\mu\nu}
F_{\alpha\beta} D_\alpha \xi_\beta{}^j ] \nonumber \\
\delta_{12} A_\mu &=& a_0 \varepsilon_{ij} D_\alpha
h_{\beta\mu}{}^i D_{[\alpha} \xi_{\beta]}{}^j
\end{eqnarray}
Here $a_0$ --- parameter having dimension $1/m^2$. But now we have
vector $B_\mu{}^i$ and scalar $\varphi^i$ as well, so we have to
consider possible non-minimal terms containing these fields too. We
have found two possible corrections for three derivatives vertex. One
of them contains tensor $h_{\mu\nu}{}^i$ and scalar $\varphi^i$ fields
with the Lagrangian:
$$
\Delta_1 {\cal L}_{13} = b_0 \varepsilon^{ij} F^{\mu\nu} [ 2 D_\mu
h_{\nu\alpha}{}^i D_\alpha \varphi^j + (D h)_\mu{}^i D_\nu \varphi^j -
D_\mu h^i D_\nu \varphi^j ]
$$
with non-trivial corrections to gauge transformations:
$$
\Delta \delta_{12} A_\mu = b_0 \varepsilon^{ij} [ D_\mu \xi_\nu{}^i
- D_\nu \xi_\mu{}^i ] D_\nu \varphi^j, \qquad
\delta_{12} \varphi^i = \frac{b_0 (d-2)}{4(d-1)} \varepsilon^{ij}
F^{\mu\nu} D_\mu \xi_\nu{}^j
$$
The other one is constructed out of gauge invariant field strengths
and does not require any corrections to gauge transformations:
$$
\Delta_2 {\cal L}_{13} = \frac{c_0}{2} \varepsilon^{ij} F^{\mu\nu} 
B_{\mu\alpha}{}^i B_{\alpha\nu}{}^j
$$
Here both $b_0$ and $c_0$ --- parameters having dimension $1/m^2$.

Now we repeat all calculations starting with the highest derivatives
terms. The structure of three derivatives vertex and two derivatives
gauge transformations is already adjusted so that all variations with
four derivatives cancel, but due to non-commutativity of $(A)dS$
covariant derivatives they give terms with two derivatives
\begin{eqnarray}
\delta_{01} {\cal L}_{13} + \delta_{12} {\cal L}_{02} &=& 2 a_0 \kappa
\varepsilon^{ij} [ (d-4) F^{\mu\nu} D_\mu h_{\nu\alpha}{}^i + (d-3)
((D h)_\mu{}^i - D_\mu h^i) F^{\mu\alpha} ] \xi_\alpha{}^j + \nonumber
\\
 && + 2 b_0 \kappa (d-3) D_\mu \varphi^i F^{\mu\nu} \xi_\nu{}^j
\label{res2}
\end{eqnarray}
which we have to take into account later. Then we add to Lagrangian
terms with two derivatives (\ref{lag2}) and corrections to gauge
transformations (\ref{cor2}) with one derivative and calculate all
variations with three derivatives. First of all their cancellation
requires that three parameters $a_0$, $b_0$ and $c_0$ satisfy a
relation:
$$
m [ a_0 (d-4) - c_0 (d-2) ] + M b_0 (d-2) = 0
$$
Once again we face an ambiguity between flat space limit and massless
limit. Indeed, in a flat space ($\kappa = 0$, $M = m$) we obtain
$b_0 = - a_0(d-4)/(d-2) + c_0$ for any mass value $m$, while for the
non-zero cosmological term $b_0 \to 0$ for $m \to 0$. We have
explicitly checked that solution exists for arbitrary values of $a_0$
and $c_0$, but in what follows we consider unambiguous case $b_0 = 0$,
$c_0 = \frac{d-4}{d-2} a_0$ only. In this, all variations with three
derivatives cancel provided:
$$
b_1 = - \frac{2 a_0 m (d-4)}{d-2}, \qquad
b_2 = - b_3 = -  \frac{2 a_0 m (d-3)}{d-2},
$$
$$
b_4 = - \frac{\beta_0}{2}, \qquad
b_5 = - b_6 = - 2 \beta_0
$$
$$
\beta_1 = - \beta_2 = 2 \beta_0, \qquad
\beta_4 = - \beta_5 = - \frac{2 a_0 m}{d-2}
$$
but again non-commutativity of $(A)dS$ covariant derivatives leaves us
with:
\begin{equation}
\delta_{01} {\cal L}_{12} + \delta_{00} {\cal L}_{13} + \delta_{11}
{\cal L}_{02} + \delta_{12} {\cal L}_{01} = 
\frac{4 a_0 \kappa m (d^2 - 5d + 7)}{d-2} \varepsilon^{ij}
B_\mu{}^i F^{\mu\nu} \xi_\nu{}^j \label{res1}
\end{equation}
At last we add to the Lagrangian terms (\ref{lag1}) with one
derivative and corrections to gauge transformations (\ref{cor1})
without derivatives and calculate all variations with two and one
derivative taking into account terms (\ref{res0}) coming from
introduction of minimal e/m interaction and terms (\ref{res2}) and
(\ref{res1}) due to non-commutativity of covariant derivatives. Their
cancellation allows us to express all parameters in terms of two main
one: $a_0$ and $\beta_0$. We obtain:
$$
a_1 = - \frac{a_0 M^2}{2(d-2)} - \frac{\beta_0 m}{2}
$$
$$
a_2 = - \frac{2 a_0 [ m^2 (d-3) + \kappa]}{d-2} - 
\frac{2 \beta_0 [ m^2 - \kappa (d-1)]}{m}
$$
$$
\alpha_1 = \frac{2 a_0 M^2}{d-2} + 2 \beta_0 m, \qquad
\alpha_3 = - \frac{4 \beta_0 M (d-1)}{d-2}
$$
$$
\alpha_4 = - \frac{4 a_0 M^2}{(d-2)^2} + 
\frac{4 \beta_0 [m^2 - \kappa (d-1)]}{m}
$$
$$
b_8 = - \frac{a_0 M (d-1)(d-4)}{(d-2)^2} - 
\frac{2 \beta_0 M (d-1)}{m(d-2)}
$$
$$
e_0 = - \frac{a_0 M^2 (d-3)}{d-2} - \beta_0 m
$$
We see that all parameters get independent additive contributions from
our two main parameters $a_0$ and $\beta_0$. As a result for any
non-zero value of $\beta_0$ part of the parameters turn out to be
singular in the massless limit. The only model that has non-singular
massless as well as flat space limits is the one with $\beta_0 = 0$.
The complete cubic vertex for such model has the form:
\begin{eqnarray}
{\cal L}_1 &=& a_0 \varepsilon_{ij} F^{\mu\nu} \left[ - D_\mu 
h_{\alpha\beta}{}^i D_\alpha h_{\beta\nu}{}^j - \frac{1}{2}
D_\alpha h_{\beta\mu}{}^i D_\alpha h_{\beta\nu}{}^j
+ D_\alpha h_{\beta\mu}{}^i D_\beta h_{\alpha\nu}{}^j +
\right. \nonumber \\
 && \qquad \qquad + \frac{1}{2} D_\mu h_{\alpha\beta}{}^i
D_\nu h_{\alpha\beta}{}^j - D_\mu h_{\nu\alpha}{}^i
(D h)_\alpha{}^j -  \frac{1}{2} (D h)_\mu{}^i (D h)_\nu{}^j +
\nonumber \\
 && \qquad \qquad + (D h)_\mu{}^i D_\nu h^j +
D_\mu h_{\nu\alpha}{}^i D_\alpha h^j - \frac{1}{2}
D_\mu h^i D_\nu h^j  +  \frac{d-4}{2(d-2)}  
B_{\mu\alpha}{}^i B_{\alpha\nu}{}^j - \nonumber \\
 && \qquad \qquad - \frac{2m}{d-2} [ (d-4) D_\mu h_{\nu\alpha}{}^i
B_\alpha{}^j + (d-3) (D h)_\mu{}^i B_\nu{}^j - (d-3) D_\mu h^i
B_\nu{}^j ] - \nonumber \\
 && \qquad \qquad - \frac{M (d-1)(d-4)}{(d-2)^2} B_{\mu\nu}{}^i
\varphi^j - \frac{M^2}{2(d-2)} h_{\mu\alpha}{}^i
h_{\nu\alpha}{}^j - \nonumber \\
 && \qquad \qquad \left. - \frac{2 [m^2 (d-3) + \kappa]}{d-2}
B_\mu{}^i B_\nu{}^j \right]
\end{eqnarray}
while non-minimal corrections to gauge transformations look like:
\begin{eqnarray}
\delta_1 A_\mu &=& a_0 \varepsilon_{ij} [ D_\alpha
h_{\beta\mu}{}^i D_{[\alpha} \xi_{\beta]}{}^j  + \frac{2m}{d-2}
B_\nu{}^i D_{[\mu} \xi_{\nu]}{}^j + \nonumber \\
 && \qquad  + \frac{2 M^2}{d-2} h_{\mu\nu}{}^i \xi_\nu{}^j -
\frac{4 M^2}{(d-2)^2} B_\mu{}^i \lambda^j ] \\
\delta_1 h_{\mu\nu}{}^i &=& a_0 \varepsilon^{ij} [ \frac{1}{2} (
F_{\mu\alpha} D_{[\alpha} \xi_{\nu]}{}^j + F_{\nu\alpha} 
D_{[\alpha} \xi_{\mu]}{}^j ) + \frac{1}{d-2} g_{\mu\nu}
F_{\alpha\beta} D_\alpha \xi_\beta{}^j ] \nonumber
\end{eqnarray}
Thus this model is a straightforward and relatively simple
generalization of our model \cite{Zin08a} for the massless particle in
$(A)dS$ space for the case of non-zero mass $m$, in this the same
cubic three derivatives vertex plays the main role. Note also that in
such model effective electric charge $e_0 = - a_0 [m^2 - \kappa (d-2)]
(d-3)/(d-2)$ becomes equal to zero exactly at the boundary of unitary
allowed region.

\section{Frame-like formalism}

In this section we reproduce the results of previous one using
frame-like gauge invariant formulation for massive spin 2 particle in
$(A)dS$ space \cite{Zin03a,Zin08b}. Such formulation being elegant and
geometric could be more compact and more suggestive for possible
generalizations on higher spins.

For the frame-like gauge invariant description of massive spin 2
particle one needs three pairs of physical and auxiliary fields:
($\omega_\mu{}^{ab}$, $h_\mu{}^a$), ($C^{ab}$, $B_\mu$) and 
($\pi^a$, $\varphi$). The Lagrangian for the free massive spin 2
particle has the form:
\begin{eqnarray}
{\cal L}_0 &=& \frac{1}{2} \left\{ \phantom{|}^{\mu\nu}_{ab} \right\}
\omega_\mu{}^{ac} \omega_\nu{}^{bc} - \frac{1}{2} \left\{
\phantom{|}^{\mu\nu\alpha}_{abc} \right\} \omega_\mu{}^{ab}
D_\nu h_\alpha{}^c + \frac{1}{8} C_{ab}{}^2 - \frac{1}{4} \left\{
\phantom{|}^{\mu\nu}_{ab} \right\} C^{ab} D_\mu B_\nu - \nonumber \\
 && -  \frac{d-1}{2(d-2)} \pi_a{}^2 + \frac{d-1}{d-2} \left\{
\phantom{|}^{\mu}_{a} \right\} \pi^a D_\mu \varphi  \nonumber \\
 && + \frac{m}{2} [ \left\{ \phantom{|}^{\mu\nu}_{ab} \right\}
\omega_\mu{}^{ab} B_\nu +  \left\{ \phantom{|}^\mu_a \right\}
C^{ab} h_\mu{}^b ] - \frac{d-1}{d-2} M \left\{ \phantom{|}^\mu_a
\right\} \pi^a B_\mu + \nonumber \\
 && + \frac{M^2}{2} \left\{ \phantom{|}^{\mu\nu}_{ab}
\right\} h_\mu{}^a h_\nu{}^b - \frac{d-1}{d-2} m M \left\{
\phantom{|}^\mu_a \right\} h_\mu{}^a \varphi + \frac{d(d-1)}{2(d-2)^2}
m^2 \varphi^2 
\end{eqnarray}
where $\left\{ \phantom{|}^{\mu\nu}_{ab} \right\} = e^\mu{}_a
e^\nu{}_b - e^\mu{}_b e^\nu{}_a$ and so on, being invariant under the
following set of initial gauge transformations:
\begin{eqnarray}
\delta_0 h_\mu{}^a &=& D_\mu \xi^a + \frac{m}{d-2} e_\mu{}^a
\xi, \qquad \delta_0 \omega_\mu{}^{ab} = D_\mu \eta^{ab} -
\frac{M^2}{d-2} e_\mu{}^{[a} \xi^{b]} \nonumber \\
\delta_0 B_\mu &=& D_\mu \xi + m \xi_\mu, \qquad \delta_0
C^{ab} = - 2m \eta^{ab}, \\
\delta_0 \varphi &=& M \xi, \qquad \delta_0 \pi^a = - m M \xi^a
\nonumber
\end{eqnarray}
where $M^2 = m^2 - \kappa (d-2)$.

Using frame-like formulation in linear approximation one is always
face an ambiguity related to the fact that there are terms in the
Lagrangian and gauge transformations which differ by terms
proportional to algebraic equations for auxiliary fields
($\omega_\mu{}^{ab}$, $C^{ab}$ and $\pi^a$ for the case at hands). Any
such Lagrangians are equivalent in this approximation but if one goes
beyond linear level things could be more complicated or simpler
depending on the choice made. In what follows we will use a kind of
$1\frac{1}{2}$ order formalism very well known from supergravity.
Namely, we will not consider any corrections to gauge transformations
for auxiliary fields $\omega_\mu{}^{ab}$, $C^{ab}$ and $\pi^a$
(usually they have the most complicated form),
instead we will require that all variations in the linear
approximation cancel up to  the terms
proportional to their free algebraic equations only. The solutions of
these free equations have the form:
$$
C_{ab} = D_{[a} B_{b]} - m h_{[ab]}, \qquad
\pi_a = D_a \varphi - M B_a
$$
$$
\omega_{a,bc} = \frac{1}{2} [ T_{ab,c} - T_{ac,b} - T_{bc,a} ] -
\frac{m}{d-2} [ g_{ab} B_c - g_{ac} B_b ]
$$
where $T_{\mu\nu}{}^a = D_\mu h_\nu{}^a - D_\nu h_\mu{}^a$. Using
these solutions one can easily derive a number of identities which
will be useful in what follows:
$$
D_{[a} \pi_{b]} = - M C_{ab} - m M h_{[ab]}, \qquad
D_{[a} C_{bc]} = - 2m D_{[a} h_{bc]}
$$
$$
R_{ab,cd} - R_{cd,ab} = \frac{m}{d-2} [ g_{ac} C_{bd} - \dots ]
+ \frac{M^2}{d-2} [ g_{ac} h_{[bd]} - \dots ]
$$
$$
R_{ab} - R_{ba} = m C_{ab} + M^2 h_{[ab]}
$$
Here $R_{\mu\nu}{}^{ab} = D_\mu \omega_\nu{}^{ab} - D_\nu
\omega_\mu{}^{ab}$, $R_{ab} = R_{ac,b}{}^c$,
while dots denote antisymmetrization on $ab$ and
$cd$.

Let us turn to the electromagnetic interactions. Here we also will
work with real fields assuming that all of them are doublets now.
First of all we introduce minimal interaction replacing all $(A)dS$
covariant derivatives in the Lagrangian and gauge transformations by
fully covariant ones, e.g.:
$$
D_\mu \xi_a{}^i \quad \Rightarrow \quad D_\mu \xi_a{}^i - e_0
\varepsilon^{ij} A_\mu \xi_a{}^j
$$
As usual such replacement spoils the invariance of the Lagrangian
under gauge transformations:
\begin{eqnarray}
\delta {\cal L}_0 &=& \frac{e_0}{2} \varepsilon^{ij} \left\{
\phantom{|}^{\mu}_{a} \right\} [ \omega_\mu{}^{bci} F^{bc} \xi^{aj} +
2 \omega_\mu{}^{abi} F^{bc} \xi^{cj} + h_\mu{}^{ai} F^{bc} \eta^{bcj}
+ 2 h_\mu{}^{bi} F^{bc} \eta^{caj} ] + \nonumber \\
 && + \frac{e_0}{2} \varepsilon^{ij} F^{ab} C_{ab}{}^i \xi^j
\label{resf0}
\end{eqnarray}

Then we proceed reconstructing three derivatives vertices in a
frame-like formalism. The main vertex that plays crucial role for
massless particle can now be written as follows \cite{Zin08a}:
\begin{equation}
{\cal L}_{13} = \frac{a_0}{8}  \varepsilon^{ij} 
\left\{ \phantom{|}^{\mu\nu}_{ab} \right\} [ 
\omega_\mu{}^{abi} F^{cd} \omega_\nu{}^{cdj}
+ \omega_\mu{}^{cdi} F^{cd} \omega_\nu{}^{abj}
- 4 \omega_\mu{}^{aci} F^{cd} \omega_\nu{}^{bdj}
- \omega_\mu{}^{cdi} F^{ab} \omega_\nu{}^{cdj} ] \label{lagf3}
\end{equation}
while appropriate corrections for gauge transformations have the form:
\begin{eqnarray}
\delta_1 h_\mu{}^{ai} &=& - \frac{a_0}{2} \varepsilon^{ij} [ F_\mu{}^b
\eta^{baj} + \eta_\mu{}^{bj} F^{ba} + \frac{1}{d-2} e_\mu{}^a 
(F \eta)^j ] \nonumber \\
\delta_1 A_\mu &=& \frac{a_0}{2} \varepsilon^{ij} \omega_\mu{}^{abi}
\eta^{abj} \label{corf3}
\end{eqnarray}
In this, the structure of the Lagrangian and gauge transformations is
already adjusted so that variations with highest number of derivatives
cancel, but due to non-commutativity of $(A)dS$ covariant derivatives
we obtain:
\begin{eqnarray}
&& \frac{a_0}{2} \varepsilon^{ij} [ - \frac{1}{2} F^{ab} 
(R_{ab,cd}{}^i - R_{cd,ab}{}^i)
\eta^{cdj} + F^{ac} (R_{ab}{}^i -
R_{ba}{}^i) \eta^{bcj} ] = \nonumber \\
&& = \frac{d-4}{2(d-2)} a_0 \varepsilon^{ij} [ m C_{ab}{}^i F^{ac}
\eta^{bcj} + M^2 h_{[ab]}{}^i F^{ac} \eta^{bcj} ] \label{resf1}
\end{eqnarray}
where in the second line we have used identities given above.

Analogously, for the second three derivatives vertex frame-like
Lagrangian and corrections to gauge transformations become:
\begin{equation}
\Delta_1 {\cal L}_{13} = b_0 \varepsilon^{ij}  \left\{
\phantom{|}^{\mu}_{a} \right\} [ 2 \omega_\mu{}^{abi} F^{bc} \pi^{cj}
- 2 \omega_\mu{}^{bci} F^{ab} \pi^{cj }+  \omega_\mu{}^{bci} F^{bc}
\pi^{aj} ]
\end{equation}
\begin{equation}
\Delta \delta_1 A_\mu = - 2 b_0 \varepsilon^{ij} \eta_\mu{}^{ai}
\pi^{aj}, \qquad \delta_1 \varphi^i = \frac{b_0(d-2)}{d-1}
\varepsilon^{ij} (F \eta)^j
\end{equation}
while the non-invariance related with non-commutativity of $(A)dS$
covariant derivatives looks like:
\begin{equation}
- 2 b_0 \varepsilon^{ij} \eta^{abi} F^{bc} D_{[a} \pi_{c]}{}^j =
2 b_0 M \varepsilon^{ij} \eta^{abi} F^{bc} [ C_{ac}{}^j + m
 h_{[ac]}{}^j ] \label{resf2}
\end{equation}
where we again used identities given above. At last the third vertex
takes the form:
\begin{equation}
\Delta_2 {\cal L}_{13} = c_0 \varepsilon^{ij} F^{ab} C_{ac}{}^i 
C_{bc}{}^j
\end{equation}

Now we turn to reformulation of two derivatives vertex. The most
general Lagrangian can be written as follows:
\begin{eqnarray}
{\cal L}_2 &=& \varepsilon^{ij} \left\{ \phantom{|}^{\mu}_{a} \right\}
[ a_1 h_\mu{}^{ai} F^{bc} C^{bcj} + a_2 h_\mu{}^{bi} F^{ac} C^{bcj} +
a_3 h_\mu{}^{bi} F^{bc} C^{acj} ] + \nonumber \\
 && + a_4 \varepsilon^{ij} \left\{ \phantom{|}^{\mu\nu}_{ab}
\right\} \omega_\mu{}^{aci} F^{bc} B_\nu{}^j + a_5 \varepsilon^{ij}
\varphi^i F^{ab} C_{ab}{}^j
\end{eqnarray}
while the most general form of corrections to gauge transformations
looks like:
\begin{equation}
\delta_1 A_\mu = \varepsilon^{ij} [ \alpha_1 \eta_\mu{}^{ai} B_a{}^j +
\alpha_2 C_{\mu a}{}^i \xi^{aj} ], \qquad
\delta_1 B_\mu{}^i = \alpha_3 \varepsilon^{ij} F_{\mu a} \xi^{aj}
\end{equation}
First of all we calculate variations under the $\xi$-transformations
and require their cancellation. This gives:
$$
a_2 = a_3 = - 2 a_1, \qquad
\alpha_2 = - 2 a_1, \qquad \alpha_3 = - 4 a_1
$$
and leaves us with non-invariance of the form:
\begin{equation}
a_1 \varepsilon^{ij} \partial_a C_{bc}{}^i [ F^{bc} \xi^{aj} - 2
F^{ac} \xi^{bj} ] = - 2m a_1 \varepsilon^{ij} \left\{
\phantom{|}^\mu_a \right\} \omega_\mu{}^{bci} [ 2 F^{ab} \xi^{cj} +
F^{bc} \xi^{aj} ] \label{resf3}
\end{equation}
where we again used identities given above. As for the invariance
under the $\eta$-transformations it requires firstly:
$$
a_4 = 0, \qquad \alpha_1 = 0
$$
It could seems that the absence of $a_4$ term in the Lagrangian and
$\alpha_1$ term in the gauge transformations contradicts with
metric-like formulation of previous section. But in the transition
from frame-like to metric-like formalism one has to solve algebraic
equations for the auxiliary fields, e.g. for $\omega$ field we get
$\omega \sim D h \oplus m B$, so that appropriate terms are already
contained in (\ref{lagf3}) and (\ref{corf3}). Secondly, we again
obtain a relation
$$
\frac{d-4}{d-2} m a_0 + 8 m c_0 = 4 M b_0
$$
having an ambiguity between flat space and massless limits. As in the
previous section, in what follows we restrict ourselves with the
unambiguous case $b_0 = 0$, $c_0 = - \frac{d-4}{8(d-2)} a_0$ only.

At last we add to the Lagrangian the most general terms with one
derivative:
\begin{equation}
{\cal L}_1 = \varepsilon^{ij} \left\{ \phantom{|}^{\mu\nu}_{ab}
\right\} [ c_1 h_\mu{}^{ci} F^{ab} h_\nu{}^{cj} + c_2 B_\mu{}^i F^{ab}
B_\nu{}^j ] + c_3 \varepsilon^{ij}  \left\{ \phantom{|}^{\mu}_{a}
\right\} h_\mu{}^{bi} F^{ab} \varphi^j
\end{equation}
as well as corresponding corrections to gauge transformations:
\begin{equation}
\delta A_\mu = \varepsilon^{ij} [ \beta_1 h_\mu{}^{ai} \xi^{aj} +
\beta_2 \varphi^i \xi_\mu{}^j + \beta_3 B_\mu{}^i \xi^j ]
\end{equation}
Then we calculate all remaining variations taking into account all
terms in (\ref{resf0}), (\ref{resf1}), (\ref{resf2}) and
(\ref{resf3}). This gives: 
$$
e_0 = - \frac{a_0 M^2 (d-3)}{d-2} + 2 m a_1, \qquad
c_1 = - \frac{a_0 M^2}{4(d-2)} + m a_1, \qquad
c_3 = \frac{4(d-1)}{d-2} M a_1
$$
$$
c_2 = \frac{a_0 M^2}{2(d-2)^2} + \frac{m^2 - \kappa (d-1)}{m}
a_1, \qquad
a_5 = - \frac{a_0 M (d-1)(d-4)}{4(d-2)^2} + 
\frac{2(d-1) M a_1}{m(d-2)}
$$
$$
\beta_1 = - 4 c_1, \qquad \beta_2 = c_3, \qquad \beta_3 = - 4 c_2
$$
Exactly as in the previous section we see that all parameters get
independent additive contributions from two main parameters $a_0$ and
$a_1$. In this, for part of the parameters contribution from $a_1$
turns out to be singular in the massless limit. So we choose simplest
case $a_1 = 0$ admitting non-singular limit $e_0 \to 0$, $m \to 0$,
$e_0/M^2 = const$. In this case complete non-minimal vertex has the
form:
\begin{eqnarray}
{\cal L}_1 &=& \frac{a_0}{8}  \varepsilon^{ij} 
\left\{ \phantom{|}^{\mu\nu}_{ab} \right\} [ 
\omega_\mu{}^{abi} F^{cd} \omega_\nu{}^{cdj}
+ \omega_\mu{}^{cdi} F^{cd} \omega_\nu{}^{abj}
- 4 \omega_\mu{}^{aci} F^{cd} \omega_\nu{}^{bdj}
- \omega_\mu{}^{cdi} F{ab} \omega_\nu{}^{cdj} ] - \nonumber \\
 && - \frac{d-4}{8(d-2)} a_0 \varepsilon^{ij} F^{ab} C_{ac}{}^i 
C_{bc}{}^j - \frac{(d-1)(d-4)}{4(d-2)^2} a_0 M
\varepsilon^{ij} \varphi^i F^{ab} C_{ab}{}^j + \nonumber \\
 && - \frac{a_0 M^2}{4(d-2)} \varepsilon^{ij} \left\{
\phantom{|}^{\mu\nu}_{ab} \right\} [ h_\mu{}^{ci} F^{ab} h_\nu{}^{cj}
- \frac{2}{d-2}  B_\mu{}^i F^{ab} B_\nu{}^j ] 
\end{eqnarray}
while corresponding corrections to gauge transformations look like:
\begin{eqnarray}
\delta_1 h_\mu{}^{ai} &=& - \frac{a_0}{2} \varepsilon^{ij} [ F_\mu{}^b
\eta^{baj} + \eta_\mu{}^{bj} F^{ba} + \frac{1}{d-2} e_\mu{}^a 
(F \eta)^j ] \nonumber \\
\delta_1 A_\mu &=& \frac{a_0}{2} \varepsilon^{ij} \omega_\mu{}^{abi}
\eta^{abj} + \frac{a_0 M^2}{d-2} \varepsilon^{ij} [ h_\mu{}^{ai}
\xi^{aj} - \frac{2}{d-2} B_\mu{}^i \xi^j ]
\end{eqnarray}
Recall that in this case effective electric charge is
$$
e_0 = - \frac{d-3}{d-2} a_0 [ m^2 - \kappa (d-2) ]
$$
so that it becomes equal to zero at the boundary of unitary allowed
region $m^2 = \kappa (d-2)$.

\section*{Conclusion and discussion}

We have shown that for massive spin 2 particles in $(A)dS$ space with
arbitrary cosmological constant it is indeed possible (at least in the
linear approximation) to switch on minimal electromagnetic
interactions supplemented by non-minimal ones containing up to three
derivatives together with corresponding non-minimal corrections to
gauge transformations. We use gauge invariant formulation for such
massive particles which works equally well both in flat Minkowski
space as well as in $(A)dS$ spaces. This allows us to construct a
model that smoothly interpolates between massless particle in $(A)dS$
space \cite{Zin08a} and massive one in flat Minkowski space.
Indeed, the relation $e_0 \sim a_0 [ m^2 - \kappa (d-2)]$ clearly
shows that having electric charge $e_0$ and our main parameter $a_0$
fixed we could easily obtain both massless limit $m \to 0$ as well as
flat limit $\kappa \to 0$, in this no singularities arise.
 Recall that both in a simple illustrative example for massive spin
3/2 particle and in our main results for massive spin 2 particle the
relations between electric charge, mass and cosmological term are such
that electric charge becomes equal to zero at the boundary of unitary
allowed regions in de Sitter space. It will be very interesting to
understand whether it is a general feature or just peculiarity of
lower spin cases.

In this paper we restrict ourselves with the linear approximation,
i.e. with the cubic vertices in the Lagrangian and linear in fields
(hence the name) corrections to gauge transformations. Let us stress
that results in this approximation do not depend on the presence of
any other fields in  the system so they are truly model independent.
If one goes beyond linear approximation, then two types of corrections
appear. From one hand, there will be terms (both in the Lagrangian and
gauge transformations) quadratic, cubic and so on in electromagnetic
field strength. Note that in a frame-like formulation linear
approximation already contains at least part of such non-linear terms
because algebraic equations for auxiliary fields look (symbolically):
$$
(1 + F) \omega \sim D h, \qquad (1 + F) C \sim D B
$$
But the most important corrections come from the fact that there are
non-trivial transformations for the e/m field $A_\mu$ itself. This, in
turn, leads to the corrections quartic in spin 2 field and it is the
consistency of these corrections that may require introduction of
(infinitely many) other fields to construct complete consistent
theory.

It is instructive to compare our results obtained here with the
results of \cite{PR08,PR08a,PR08b}. Authors also use gauge invariant
description for massive particles, but they insist that the whole
Lagrangian has to be written in terms of gauge invariant combination
$$
\tilde{h}_{\mu\nu} = h_{\mu\nu} + \frac{1}{m} D_{(\mu} B_{\nu)} -
\frac{1}{2m^2} D_{(\mu} D_{\nu)} \phi
$$
thus completely ignoring the possibility to consider non-minimal
corrections for gauge transformations. Clearly, any Lagrangian
constructed this way will be trivially gauge invariant, but the price
is that it will contain too many derivatives. Moreover, it is clear
that such gauge invariant formulation will be equivalent to the
initial one and share all its problems. One of the well known problems
that arise when one consider massive spin 2 particles in
electromagnetic or gravitational background \cite{DW01d,BD72} is the
(re)appearance of sixth ghost degree of freedom. At the same time the
model constructed in this paper has right number of physical degrees
of freedom.

For simplicity let us consider flat $d=4$ Minkowski space and choose a
unitary gauge $B_\mu = 0$, $\varphi = 0$. In this gauge the model
looks like the usual non-gauge invariant theory for massive spin
particle with the free Lagrangian:
\begin{eqnarray*}
{\cal L}_0 &=& \frac{1}{2} D^\alpha h^{\mu\nu} D_\alpha h_{\mu\nu}
- \frac{1}{2} D^\alpha h^{\mu\nu} D_\mu h_{\nu\alpha} - \frac{1}{2}
(D h)^\mu (D h)_\mu + (D h)^\mu D_\mu h - \frac{1}{2} D^\mu h D_\mu h
- \\
 && - \frac{m^2}{2} ( h^{\mu\nu} h_{\mu\nu} - h^2 )
\end{eqnarray*}
and linear part of non-minimal vertex having the form:
\begin{eqnarray*}
{\cal L}_1 &=& a_0 \varepsilon_{ij} F^{\mu\nu} \left[ - D_\mu 
h_{\alpha\beta}{}^i D_\alpha h_{\beta\nu}{}^j - \frac{1}{2}
D_\alpha h_{\beta\mu}{}^i D_\alpha h_{\beta\nu}{}^j
+ D_\alpha h_{\beta\mu}{}^i D_\beta h_{\alpha\nu}{}^j +
\right. \\
 && \qquad \qquad + \frac{1}{2} D_\mu h_{\alpha\beta}{}^i
D_\nu h_{\alpha\beta}{}^j - D_\mu h_{\nu\alpha}{}^i
(D h)_\alpha{}^j -  \frac{1}{2} (D h)_\mu{}^i (D h)_\nu{}^j + \\
 && \qquad \qquad \left. + (D h)_\mu{}^i D_\nu h^j +
D_\mu h_{\nu\alpha}{}^i D_\alpha h^j - \frac{1}{2}
D_\mu h^i D_\nu h^j - \frac{m^2}{4} h_{\mu\alpha}{}^i
h_{\nu\alpha}{}^j \right]
\end{eqnarray*}
By straightforward calculations we can show that all usual constraints
do follow from the equations of motion. We obtain:
\begin{eqnarray*}
&& D^\nu ( \frac{\delta {\cal L}_0}{\delta h_{\mu\nu}} +
\frac{\delta {\cal L}_1}{\delta h_{\mu\nu}}) - \frac{a_0}{4} [
2 D^\alpha F^{\beta\mu} + 2 F^{\beta\mu} D^\alpha - 2 F^{\alpha\nu}
D_\nu g^{\beta\mu} - F^{\mu\nu} D_\nu g^{\alpha\beta} ] 
\frac{\delta {\cal L}_0}{\delta h_{\alpha\beta}} = \\
&& = - m^2 (g^{\mu\nu} - \frac{a_0}{2} F^{\mu\nu})
 ((D h)_\nu - D_\nu h) = 0 \\
&& (D^\mu D^\nu - \frac{m^2}{2}) ( \frac{\delta {\cal L}_0}{\delta
h_{\mu\nu}} + \frac{\delta {\cal L}_1}{\delta h_{\mu\nu}}) =
- \frac{3m^4}{2} h = 0
\end{eqnarray*}
where we have omitted all terms quadratic in $F_{\mu\nu}$ as well as
terms proportional to free e/m equation $(D F)_\mu = 0$ as it is
appropriate for linear approximation. Thus, though gauge invariance
for massive particles does not automatically guarantee the right
number of physical degrees of freedom, it indeed can help to construct
such models. To our opinion, the right way is to consider not only the
most general non-minimal higher derivatives interactions, but also the
most general non-minimal corrections to gauge transformations with the
additional requirement that algebra of gauge transformations closes.
In this, the best strategy is to use minimal number of derivatives
possible and avoid trivial solutions related with the substitution
$h_{\mu\nu} \rightarrow \tilde{h}_{\mu\nu}$. It is interesting to note
that the same cubic vertex
$$
{\cal L} \sim D h F D h
$$
may be needed for the theory to be causal \cite{AN89}. 

Now let us turn to the behavior of tree amplitudes at high energies.
In non-gauge invariant description one has to work with the usual
propagator for massive spin 2 particle which has the terms up to
$p^\mu p^\nu p^\alpha p^\beta/m^4$ leading, in general, to a very bad
high energy behavior. But for some very specific combinations of
non-minimal terms one face a number of cancellations. They happen each
time when divergency $D^\mu J_{\mu\nu}$ or double divergency 
$D^\mu D^\nu J_{\mu\nu}$ of the "current" (for spin 2 it also is a
second rank tensor) turn out to be proportional to free equations of
motion because external legs are on shell. Thus to obtain the correct
high energy behavior one have to make careful calculations with the
full propagator and appropriate vertices. But as authors of 
\cite{PR08,PR08a,PR08b} teach us, there is a more simple and elegant
way. Let us introduce auxiliary fields $B_\mu$ and $\varphi$ and make
our Lagrangian to be gauge invariant. Then working perturbatively we
can always choose the gauge (analog of so-called renormalizable gauge
for spontaneously broken Yang-Mills theories) where free Lagrangian is
diagonal and we have three independent components $h_{\mu\nu}$,
$B_\mu$ and $\varphi$ all with the nice propagators $1/(p^2 - m^2)$.
Thus the behavior of amplitudes can be easily extracted right from the
interacting Lagrangian expressed in terms of these fields. But as we
have seen in this paper there are may be different ways to make the
same initial Lagrangian to be gauge invariant and, as a result, such
estimates may be drastically different.

Let us take the model constructed here and consider simple
tree level diagrams like the scattering of massive spin 2 particles
due to one-photon exchange or Compton scattering. 
Then from the cubic vertices we will obtain for both amplitudes
$e p^3 (1/p^2) e p^3 \rightarrow e^2 p^4$, where $e p^3$ comes from
the most hard vertex $e Dh F Dh$ and $1/p^2$ comes from propagators.
But we have to take into account that both amplitudes will gain
contributions from quartic vertices. The explicit structure of such
vertices crucially depends on the presence or absence of other fields
in the system. Let us suppose that we will not introduce any other
fields (really it is a worst case because no wonderful cancellations
can happen). Then having Argyres-Nappi vertex $(e/m^2) Dh F Dh$ and
gauge transformation of the form $\delta h \sim F D \xi$ and 
$\delta A \sim Dh D \xi$, we will have to introduce at the quadratic
approximation two type of quartic vertices. Firstly, we will have
second Argyres-Nappi vertex $(e^2/m^4) Dh F^2 Dh$ which will produce
the same $e^2 p^4$ contribution to Compton scattering. Secondly, we
will get quartic vertex of the form $(e^2/m^4) (Dh)^4$ which will
again produce the same $e^2 p^4$ contribution to one-photon exchange.
Thus (leaving aside a tiny probability of some wonderful cancellation)
we expect that both amplitudes will behave like $e^2 p^4$ at high
energies. And it seems very natural that it is the model with the
right number of physical degrees of freedom and without any
ghosts that has the best high energy behavior (compare e.g.
\cite{DR05,CNPT05}).


\begin{thebibliography}{10}

\bibitem{AD79}
C.~Aragone, S.~Deser
{\it "Consistency problem of hypergravity",}
Phys. Lett. {\bf B86} (1979) 161.

\bibitem{WF80}
D.~de~Wit, D.~Z. Freedman
{\it "Systematics of higher-spin gauge fields",}
Phys. Rev. {\bf D21} (1980) 358.

\bibitem{BBD85}
F.~A. Berends, G.~J.~H. Bugrers, H.~van Dam
{\it "On the theoretical problems in constructing interactions
involving higher-spin massless particles",}
Nucl. Phys. {\bf B260} (1985) 295.

\bibitem{Por08}
M.~Porrati
{\it "Universal Limits on Massless High-Spin Particles",}
arXiv:0804.4672.

\bibitem{FV87}
E.~S. Fradkin, M.~A. Vasiliev
{\it "On the gravitational interaction of massless higher-spin
fields",}
Phys. Lett. {\bf B189} (1987) 89.

\bibitem{FV87a}
E.~S. Fradkin, M.~A. Vasiliev
{\it "Cubic interaction in extended theories of massless higher-spin
fields",}
Nucl. Phys. {\bf B291} (1987) 141.

\bibitem{Zin08}
Yu.~M. Zinoviev
{\it "On spin 3 interacting with gravity",}
Class. Quantum Grav. {\bf 26} (2009) 035022, arXiv:0805.2226.

\bibitem{BL06}
N.~Boulanger, S.~Leclercq
{\it "Consistent couplings between spin-2 and spin-3 massless
fields",}
JHEP {\bf 0611} (2006) 034, arXiv:hep-th/0609221.

\bibitem{BLS08}
N.~Boulanger, S.~Leclercq, P.~Sundell
{\it "On The Uniqueness of Minimal Coupling in Higher-Spin Gauge
Theory",}
JHEP {\bf 0808} (2008) 056, arXiv:0805.2764.

\bibitem{Zin08a}
Yu.~M. Zinoviev
{\it "On spin 2 electromagnetic interactions",}
Mod. Phys. Lett. {\bf A24} (2009) 17, arXiv:0806.4030.

\bibitem{OP65}
V.~I. Ogievetsky, I.~V. Polubarinov
{\it "Interacting field of spin 2 and the Einstein equations",}
Ann. Phys. {\bf 35} (1965) 167.

\bibitem{FF79}
J.~Fang, C.~Fronsdal
{\it "Deformations of gauge groups. Gravitation",}
J. Math. Phys. {\bf 20} (1979) 2264.

\bibitem{MUF80}
K.~A. Milton, L.~F. Urrutia, R.~J. Finkelstein
{\it "Constructive approach to supergravity",}
Gen. Rel. Grav. {\bf 12} (1980) 67.

\bibitem{Wald86}
R.~M. Wald
{\it "Spin-two fields and general covariance",}
Phys. Rev. {\bf D33} (1986) 3613.

\bibitem{BH93}
G.~Barnich and M.~Henneaux
{\it "Consistent couplings between fields with a gauge freedom and
deformations of the master equation",}
Phys. Lett.  {\bf B311} (1993) 123, arXiv:hep-th/9304057.

\bibitem{Hen98}
M.~Henneaux
{\it "Consistent interactions between gauge fields: The cohomological
approach",}
Contemp. Math.  {\bf 219} (1998) 93, arXiv:hep-th/9712226. 

\bibitem{BBCL06}
X.~Bekaert, N.~Boulanger, S.~Cnockaert and S.~Leclercq
{\it "On Killing tensors and cubic vertices in higher-spin gauge
theories",}
Fortsch. Phys.  {\bf 54} (2006) 282, arXiv:hep-th/0602092.

\bibitem{BFPT06}
I.~L. Buchbinder, A.~Fotopoulos, A.~C. Petkou, M.~Tsulaia
{\it "Constructing the Cubic Interaction Vertex of Higher Spin Gauge
Fields",}
Phys. Rev. {\bf D74} (2006) 105018, arXiv:hep-th/0609082.

\bibitem{HS82}
C.~R. Hagen, L.~P.~S. Singh
{\it "Search for consistent interactions of the Rarita-Schwinger
field",}
Phys. Rev. {\bf D26} (1982) 393.

\bibitem{FPT92}
S.~Ferrara, M.~Porrati, V.~L. Telegdi
{\it "$g=2$ as the natural value of the tree-level gyromagnetic ratio
of elementary particles",}
Phys. Rev. {\bf D46} (1992) 3529.

\bibitem{CPD94}
A.~Cucchieri, S.~Deser, M.~Porrati
{\it "Tree-level unitarity constraints on the gravitational couplings
of higher-spin massive fields",}
Phys. Rev. {\bf D51} (1995) 4543, arXiv:hep-th/9408073.

\bibitem{VZ69}
G.~Velo and D.~Zvanziger
{\it "Propagation and quantization of Rarita-Schwinger waves in an
external electromagnetic potential",}
Phys. Rev. {\bf D22} (1969) 1337.

\bibitem{DPW00}
S.~Deser, V.~Pascalutsa, A.~Waldron
{\it "Massive spin 3/2 electrodynamics",}
Phys. Rev. {\bf D62} (2000) 105031, arXiv:hep-th/0003011.

\bibitem{DW01d}
S.~Deser, A.~Waldron
{\it "Inconsistencies of massive charged gravitating higher spins",}
Nucl. Phys. {\bf B631} (2002) 369, arXiv:hep-th/0112182.

\bibitem{BKP99}
I.~L. Buchbinder, V.~A. Krykhtin, V.~D. Pershin
{\it "On Consistent Equations for Massive Spin-2 Field Coupled to
Gravity in String Theory",}
Phys. Lett. {\bf B466} (1999) 216, arXiv:hep-th/9908028.

\bibitem{BGKP99}
I.~L. Buchbinder, D.~M. Gitman, V.~A. Krykhtin, V.~D. Pershin
{\it "Equations of Motion for Massive Spin 2 Field Coupled to
Gravity",}
Nucl. Phys. {\bf B584} (2000) 615, arXiv:hep-th/9910188.

\bibitem{BGP00}
I.~L. Buchbinder, D.~M. Gitman, V.~D. Pershin
{\it "Causality of Massive Spin 2 Field in External Gravity",}
Phys. Lett. {\bf B492} (2000) 161, arXiv:hep-th/0006144.

\bibitem{BK05}
I.~L. Buchbinder, V.~A. Krykhtin
{\it "Gauge invariant Lagrangian construction for massive bosonic
higher spin fields in D dimensions",}
Nucl. Phys. {\bf B727} (2005) 537, arXiv:hep-th/0505092.

\bibitem{BKRT06}
I.~L. Buchbinder, V.~A. Krykhtin, L~.L. Ryskina, H.~Takata
{\it "Gauge invariant Lagrangian construction for massive higher spin
fermionic fields",}
Phys. Lett. {\bf B641} (2006) 386, arXiv:hep-th/0603212.

\bibitem{BKL06}
I.~L. Buchbinder, V.~A. Krykhtin, P.~M. Lavrov
{\it "Gauge invariant Lagrangian formulation of higher spin massive
bosonic field theory in AdS space",}
Nucl. Phys. {\bf B762} (2007) 344, arXiv:hep-th/0608005.

\bibitem{BKR07}
I.~L. Buchbinder, V.~A. Krykhtin, A.~A. Reshetnyak
{\it "BRST approach to Lagrangian construction for fermionic higher
spin fields in (A)dS space",}
Nucl. Phys. {\bf B787} (2007) 211, arXiv:hep-th/0703049.

\bibitem{MR07}
P.~Yu. Moshin, A.~A. Reshetnyak
{\it "BRST approach to Lagrangian formulation for mixed-symmetry
fermionic higher-spin fields",}
JHEP {\bf 10} (2007) 040, arXiv:0707.0386.

\bibitem{BKT07}
I.~L. Buchbinder, V.~A. Krykhtin, H.~Takata
{\it "Gauge invariant Lagrangian construction for massive bosonic
mixed symmetry higher spin fields",}
Phys. Lett. {\bf B656} (2007) 253, arXiv:0707.2181.

\bibitem{Zin83}
Yu.~M. Zinoviev
{\it "Gauge invariant description of massive high spin particles",}
Preprint 83-91, IHEP, Protvino, 1983.

\bibitem{KZ97}
S.~M. Klishevich, Yu.~M. Zinoviev
{\it "On electromagnetic interaction of massive spin-2 particle",}
Phys. Atom. Nucl. {\bf 61} (1998) 1527, arXiv:hep-th/9708150.

\bibitem{Zin01}
Yu.~M. Zinoviev
{\it "On Massive High Spin Particles in (A)dS",}
arXiv:hep-th/0108192.

\bibitem{Met06}
R.~R. Metsaev
{\it "Gauge invariant formulation of massive totally symmetric
fermionic fields in (A)dS space",}
Phys. Lett. {\bf B643} (2006) 205-212, arXiv:hep-th/0609029.

\bibitem{Zin02a}
Yu.~M. Zinoviev
{\it "On Massive Mixed Symmetry Tensor Fields in Minkowski space and
(A)dS",}
arXiv:hep-th/0211233.

\bibitem{Zin03a}
Yu.~M. Zinoviev
{\it "First Order Formalism for Massive Mixed Symmetry Tensor Fields
in Minkowski and (A)dS Spaces",}
arXiv:hep-th/0306292.

\bibitem{Med03}
P.~de~Medeiros
{\it "Massive gauge-invariant field theories on spaces of constant
curvature",}
Class. Quant. Grav. {\bf 21} (2004) 2571, arXiv:hep-th/0311254.

\bibitem{BHR05}
M.~Bianchi, P.~J. Heslop, F.~Riccioni
{\it "More on La Grande Bouffe",}
JHEP {\bf 08} (2005) 088, arXiv:hep-th/0504156.

\bibitem{HW05}
K.~Hallowell, A.~Waldron
{\it "Constant Curvature Algebras and Higher Spin Action Generating
Functions",}
Nucl. Phys. {\bf B724} (2005) 453, arXiv:hep-th/0505255.

\bibitem{BG08}
I.~L. Buchbinder, A.~V. Galajinsky
{\it "Quartet Unconstrained Formulation for Massive Higher Spin
Fields",}
JHEP {\bf 0811} (2008) 081, arXiv:0810.2852.

\bibitem{DW01}
S.~Deser, A.~Waldron
{\it "Gauge Invariance and Phases of Massive Higher Spins in (A)dS",}
Phys. Rev. Lett. {\bf 87} (2001) 031601, arXiv:hep-th/0102166.

\bibitem{DW01a}
S.~Deser, A.~Waldron
{\it "Partial Masslessness of Higher Spins in (A)dS",}
Nucl. Phys. {\bf B607} (2001) 577, arXiv:hep-th/0103198.

\bibitem{DW01c}
S.~Deser, A.~Waldron
{\it "Null Propagation of Partially Massless Higher Spins in (A)dS and
  Cosmological Constant Speculations",}
Phys. Lett. {\bf B513} (2001) 137, arXiv:hep-th/0105181.

\bibitem{SV06}
E.~D. Skvortsov, M.~A. Vasiliev
{\it "Geometric Formulation for Partially Massless Fields",}
Nucl. Phys. {\bf B756} (2006) 117, arXiv:hep-th/0601095.

\bibitem{DW06}
S.~Deser, A.~Waldron
{\it "Partially Massless Spin 2 Electrodynamics",}
Phys. Rev. {\bf D74} (2006) 084036, arXiv:hep-th/0609113.

\bibitem{Zin06}
Yu.~M. Zinoviev
{\it "On massive spin 2 interactions",}
Nucl. Phys. {\bf B770} (2007) 83-106, arXiv:hep-th/0609170.

\bibitem{Met06a}
R.~R. Metsaev
{\it "Gravitational and higher-derivative interactions of massive spin
5/2 field in (A)dS space",}
Phys. Rev. {\bf D77} (2008) 025032, arXiv:hep-th/0612279.

\bibitem{SH74}
L.~P.~S. Singh, C.~R. Hagen
{\it "Lagrangian formulation for arbitrary spin. 1. The boson case.",}
Phys. Rev. {\bf D9} (1974) 898.

\bibitem{SH74a}
L.~P.~S. Singh, C.~R. Hagen
{\it "Lagrangian formulation for arbitrary spin. 2. The fermion
case.",}
Phys. Rev. {\bf D9} (1974) 910.

\bibitem{Fro78}
C.~Fronsdal
{\it "Massless fields with integer spin",}
Phys. Rev. {\bf D18} (1978) 3624.

\bibitem{FF78}
J.~Fang, C.~Fronsdal
{\it "Massless fields with half integral spin",}
Phys. Rev. {\bf D18} (1978) 3630.

\bibitem{Fro79}
C.~Fronsdal
{\it "Singletons and massless, integral spin fields on de sitter
space",}
Phys. Rev. {\bf D20} (1979) 848.

\bibitem{FF80}
J.~Fang, C.~Fronsdal
{\it "Massless half integer spin fields in de sitter space",}
Phys. Rev. {\bf D22} (1980) 1361.

\bibitem{Vas80}
M.~A. Vasiliev
{\it "'Gauge' form of description of massless fields with arbitrary
spin",}
Sov. J. Nucl. Phys. {\bf 32} (1980) 439.

\bibitem{LV88}
V.~E. Lopatin, M.~A. Vasiliev
{\it "Free massless bosonic fields of arbitrary spin in d-dimensional
de sitter space",}
Mod. Phys. Lett. {\bf A3} (1988) 257.

\bibitem{Vas88}
M.~A. Vasiliev
{\it "Free massless fermionic fields of arbitrary spin in
d-dimensional de sitter space",}
Nucl. Phys, {\bf B301} (1988) 26.

\bibitem{AD80}
C.~Aragone, S.~Deser
{\it "Higher spin vierbein gauge fermions and hypergravities",}
Nucl. Phys. {\bf B170} (1980) 329.

\bibitem{Zin03}
Yu.~M. Zinoviev
{\it "First Order Formalism for Mixed Symmetry Tensor Fields",}
arXiv:hep-th/0304067.

\bibitem{ASV03}
K.B. Alkalaev, O.V. Shaynkman, M.A. Vasiliev
{\it "On the Frame-Like Formulation of Mixed-Symmetry Massless Fields
in (A)dS(d)",}
Nucl. Phys. {\bf B692} (2004) 363, arXiv:hep-th/0311164.

\bibitem{ASV05}
K.B. Alkalaev, O.V. Shaynkman, M.A. Vasiliev
{\it "Lagrangian Formulation for Free Mixed-Symmetry Bosonic Gauge
Fields in (A)dS(d)",}
JHEP {\bf 0508} (2005) 069, arXiv:hep-th/0501108.

\bibitem{ASV06}
K.~B. Alkalaev, O.~V. Shaynkman, M.~A. Vasiliev
{\it "Frame-like formulation for free mixed-symmetry bosonic massless
  higher-spin fields in AdS(d)",} arXiv:hep-th/0601225.

\bibitem{SV08}
D.~P. Sorokin, M.~A. Vasiliev
{\it "Reducible higher-spin multiplets in flat and AdS spaces and
their geometric frame-like formulation",} arXiv:0807.0206.

\bibitem{Skv08}
E.~D. Skvortsov
{\it "Frame-like Actions for Massless Mixed-Symmetry Fields in
Minkowski space",}
Nucl. Phys. {\bf B808} (2009) 569, arXiv:0807.0903.

\bibitem{Zin08b}
Yu.~M. Zinoviev
{\it "Frame-like gauge invariant formulation for massive high spin
particles",}
Nucl. Phys. {\bf B808} (2009) 185, arXiv:0808.1778.

\bibitem{Zin08c}
Yu.~M. Zinoviev
{\it "Towards frame-like gauge invariant formulation for massive mixed
symmetry bosonic fields",}
Nucl. Phys. {\bf B812} (2009) 46, arXiv:0809.3287.

\bibitem{BIS08}
N.~Boulanger, C.~Iazeolla, P.~Sundell
{\it "Unfolding Mixed-Symmetry Fields in AdS and the BMV Conjecture:
I. General Formalism",}
arXiv:0812.4438.

\bibitem{BIS08a}
N.~Boulanger, C.~Iazeolla, P.~Sundell
{\it "Unfolding Mixed-Symmetry Fields in AdS and the BMV Conjecture:
II. Oscillator Realization",}
arXiv:0812.4438.

\bibitem{Gar03}
T.~Garidi
{\it "What is mass in desitterian physics?",} arXiv:hep-th/0309104.

\bibitem{DW00}
S.~Deser, A.~Waldron
{\it "(Dis)continuities of Massless Limits in Spin 3/2-mediated
Interactions and Cosmological Supergravity",}
Phys. Lett. {\bf B501} (2001) 134-139, arXiv:hep-th/0012014.

\bibitem{PR08}
M.~Porrati, R.~Rahman
{\it "Intrinsic Cutoff and Acausality for Massive Spin 2 Fields
Coupled to Electromagnetism",}
Nucl. Phys. {\bf B801} (2008) 174, arXiv:0801.2581.

\bibitem{PR08a}
M.~Porrati, R.~Rahman
{\it "Electromagnetically Interacting Massive Spin-2 Field: Intrinsic
Cutoff and Pathologies in External Fields",} arXiv:0809.2807.

\bibitem{PR08b}
M.~Porrati, R.~Rahman
{\it "A Model Independent Ultraviolet Cutoff for Theories with Charged
Massive Higher Spin Fields",} arXiv:0812.4254.

\bibitem{BD72}
D.~G. Boulware, S.~Deser
{\it "Can Gravitation Have a Finite Range?"}
Phys. Rev. {\bf D6} (1972) 3368.

\bibitem{AN89}
P.~C. Argyres, C.~R. Nappi
{\it "Massive Spin-2 Bosonic String States in an Electromagnetic
Background",}
Phys. Lett. {\bf B224} (1989) 89.

\bibitem{DR05}
C.~Deffayet, J.-W. Rombouts
{\it "Ghosts, Strong Coupling and Accidental Symmetries in Massive
Gravity",}
Phys. Rev. {\bf D72} (2005) 044003, arXiv:gr-qc/0505134.

\bibitem{CNPT05}
P.~Creminelli, A.~Nicolis, M.~Papucci, E.~Trincherini
{\it "Ghosts in Massive Gravity",}
JHEP {\bf 0509} (2005) 003, arXiv:hep-th/0505147.

\end{thebibliography}
\end{document}